\def\tr{{\text{tr}}\,}

\def\be{\begin{equation}}
\def\ee{\end{equation}}
\def\bea{\begin{eqnarray}}
\def\eea{\end{eqnarray}}
\def\bse{\begin{subequations}}
\def\ese{\end{subequations}}

\documentclass[prb,twocolumn,showpacs,amsmath,amssymb,eqsecnum]{revtex4} 
\usepackage{graphicx}% Include figure files
\usepackage{dcolumn}% Align table columns on decimal point
\usepackage{bm}% bold math
\begin{document}
\title{Nature of the Quantum Phase Transition in Clean, Itinerant Heisenberg
       Ferromagnets
       %\small{$[$ Phys. Rev. B {\bm 63}, 174428 (2001) $]$
}
\author{T.R.Kirkpatrick}
\affiliation{Institute for Physical Science and Technology, and Department of 
         Physics\\
         University of Maryland, College Park, MD 20742}
\author{D.Belitz}
\affiliation{Department of Physics and Materials Science Institute, %\\
         University of Oregon, %\\
         Eugene, OR 97403}
\date{\today}

\begin{abstract}
A comprehensive theory of the quantum phase transition in clean, 
itinerant Heisenberg ferromagnets is presented. It is shown that the 
standard mean-field description of the transition is invalid in spatial 
dimensions $d\leq 3$ due to the existence of soft particle-hole excitations 
that couple to the order parameter fluctuations and lead to an upper critical
dimension $d_{\text{c}}^{\, +} = 3$. A generalized mean-field theory that 
takes these additional modes into account predicts a fluctuation-induced 
first-order transition. In a certain parameter regime, this first-order 
transition in turn is unstable with respect to a fluctuation-induced 
second-order transition. The quantum ferromagnetic
transition may thus be either of first or of second-order, in agreement
with experimental observations. A detailed discussion is given of the 
stability of the first-order transition, and of the critical behavior at
the fluctuation-induced second-order transition. In $d=3$, the latter is
mean field-like with logarithmic corrections to scaling, and in $d<3$ it
can be controlled by means of a $3-\epsilon$ expansion.
\end{abstract}


\maketitle

\section{Introduction}
\label{sec:I}

The study of quantum phase transitions (QPTs) is currently an important and
very active field of research in condensed matter physics, see, e.g.,
Ref.\ \onlinecite{Sachdev_99}. Although, strictly speaking, these transitions 
occur only at zero temperature ($T=0$), they are important
for understanding the behavior of many systems at low, but routinely 
accessible, temperatures. Understanding QPTs is also important for gaining
insight into the possible phases of systems at zero temperature. Indeed, 
QPTs are thought to be relevant for understanding phenomena as diverse
as high-$T_{\text{c}}$ superconductivity, the quantum Hall effects, various 
magnetic phenomena in both metallic and insulating systems, the transport
properties of doped semiconductors, and superconductor-metal and 
superconductor-insulator transitions, see, e.g., 
Refs.\ \onlinecite{Sondhi_et_al_97,Sarachik_01,us_R,Coleman_Maple_Millis_96}.

Perhaps the most obvious, and one might naively think almost trivial, QPT is
the ferromagnetic transition that takes place in a clean
\footnote{By ``clean'' we mean a system that is free of impurities or 
 structural disorder.}
itinerant electron system as the exchange 
coupling is varied at zero temperature. Indeed, this was one of the first 
QPTs to be considered, see Ref.\ \onlinecite{Hertz_76} and references
therein. The traditional arguments and results for 
this QPT can be paraphrased as follows. Let ${\bm M}$ be the order parameter
vector, i.e., the magnetization, with components $M^i$ ($i=x,y,z$). Landau 
theory,\cite{Landau_Lifshitz_V_80} which as a mean-field description is 
suitable for both thermal and 
quantum phase transitions, says that the free energy (at $T>0$),
or energy (at $T=0$), as a function of ${\bm M}$, for small 
magnetization and small magnetic field ${\bm h}$ is of the form
\be
F = F_0 + t\,{\bm M}^2 + u\,{\bm M}^4 
          - {\bm h}\cdot{\bm M}\quad.  
\label{eq:1.1}
\ee
Here $F_0$ is a background contribution that reflects the degrees of freedom
other than the order parameter. $t$ turns out to be the distance from the 
mean-field transition, i.e., the transition takes place at $t=0$, and $u$ is 
a constant that is assumed to be positive. Minimizing Eq.\ (\ref{eq:1.1}) 
with respect to ${\bm M}$ leads to the conclusion that at zero external 
magnetic field the magnetic transition is continuous with mean-field or
Landau critical exponents. Fluctuations invalidate Landau theory in
sufficiently low dimensions, while the mean-field critical behavior is
exact in dimensions $d$ larger then an upper critical dimension 
$d_{\text{c}}^{\, +}$. For the thermal 
phase transition in a Heisenberg ferromagnet, it is well established that
$d_{\text{c}}^{\, +} = 4$. For the corresponding QPT, it was argued that 
$d_{\text{c}}^{\, +} = 4 - z$, with $z$ the dynamical scaling 
exponent.\cite{Hertz_76} This reduction of the upper critical dimension is a 
result of the coupling between statics and dynamics in quantum statistical
mechanics, which leads to an effective dimension for fluctuations given by 
$d_{\text{eff}} = d + z$. Mean-field theory suggests $z = 3$ for the quantum
Heisenberg transition of clean itinerant electrons, so the conclusion was 
that this QPT should have a $d_{\text{c}}^{\, +} = 1$, resulting in mean 
field-like critical behavior for both thin films and bulk systems. From a 
theoretical statistical mechanics point of view, the itinerant quantum 
ferromagnetic transition therefore did not appear to be very interesting.

This conclusion was recently challenged by what amounts to a generalized 
mean-field description of the transition.\cite{us_1st_order_PRL} The basic
physical argument, which is general and applies to other phase transitions
as well, is as follows. In the disordered phase, $F_0$ 
contains contributions from fermionic soft modes, viz., particle-hole
excitations. Some of these acquire a mass in the ordered phase,
which decreases the contribution of these fluctuations to the free energy,
and thus leads to a negative term in the free energy function that has a
nonanalytic dependence on the order parameter. 
\footnote{The sign of this term is consistent with the fact that the
 mode-mode coupling contribution is due to fluctuation effects that tend to
 suppress ferromagnetism. In a system with quenched disorder, the effect
 has the opposite sign. See Ref.\ \onlinecite{us_woelfle} for a discussion
 of this point.}
If this mode-mode coupling
effect, which is neglected in the usual Landau or mean-field theory, is
strong enough, it clearly can lead to a modification of phase transition
predicted by Eq.\ (\ref{eq:1.1}). Ref.\ \onlinecite{us_1st_order_PRL} showed
that in the case of an itinerant ferromagnet, the soft modes that couple
most strongly to the order parameter, viz., spin-triplet particle-hole
excitations, do indeed develop a mass in the ordered phase and lead to a
Landau energy function that has the form, in $d=3$,
\be
F = F_0 + t\,{\bm M}^2 + v\,{\bm M}^4\ln{\bm M}^2
        + u\,{\bm M}^4 - {\bm h}\cdot{\bm M}\quad.
\label{eq:1.2}
\ee
The presence of the ${\bm M}^4\ln{\bm M}^2$ term, compared to 
Eq.\ (\ref{eq:1.1}), changes the nature of the transition from 
a continuous one to a discontinuous one. The same is true for dimensions 
$1<d<3$, due to a similar nonanalytic term in the Landau function. 
The fact that the nature of the 
phase transition in $d=3$ changes qualitatively upon improving on Landau 
theory is not consistent with the traditional notion of a $3$-$d$ system 
being above its upper critical dimension $d_{\text{c}}^{\, +} = 1$. In
contrast to the traditional prediction of a continuous transition with
mean-field exponents, the generalized mean-field theory predicts the 
transition to {\em always} be of first order provided that $d\leq 3$.

Experimentally, the situation is seemingly inconclusive. In some ferromagnets
with low Curie temperatures, where the quantum phase transition can be
triggered by hydrostatic pressure, or composition, the observed transition
is of first-order, in agreement with the generalized mean-field theory. This
is the case, for instance, in MnSi,\cite{Pfleiderer_et_al_97} and in 
UGe$_2$.\cite{Saxena_et_al_00} In others, for instance,
ZrZn$_2$,\cite{Pfleiderer_et_al_97} and 
Ni$_{\text{x}}$Pd$_{1-{\text{x}}}$,\cite{Nicklas_et_al_99} however, 
the transition is observed to be continuous. Moreover, the critical behavior 
observed in Ni$_{\text{x}}$Pd$_{1-{\text{x}}}$ is in good agreement with 
mean-field exponents. This is surprising, given the above conclusion that 
mean-field theory cannot be correct in $d=3$.

In this paper we provide new insights into this QPT, and additional
understanding of the discontinuous transition that results from the
generalized mean-field theory, as well as of the stability of the latter.
In the general theory of phase transitions, transitions that are predicted 
to be continuous by Landau theory but are in fact discontinuous are called 
fluctuation-induced first-order phase 
transitions.\cite{Halperin_Lubensky_Ma_74} We will show that the 
first-order transition in itinerant quantum ferromagnets can indeed be 
understood as being fluctuation-induced. The novel feature is that although 
the order parameter fluctuations are above their upper critical dimension, 
in a well-defined sense the soft fermion fluctuations are not, and it is the 
effect of these fluctuations that drives the transition first order for 
$d\leq 3$. 

Their are many similarities between the fluctuation-induced first-order
phase transition discussed here, and the thermal first-order transition 
that occurs in conventional superconductors, or the nematic-to-smectic-A 
transition in liquid crystal 
systems.\cite{Halperin_Lubensky_Ma_74,Chen_Lubensky_Nelson_78} 
In all of these systems there are 
soft or massless excitations (in superconductors, these are the gauge or 
vector potential fluctuations; in liquid crystals, the director fluctuations; 
while in the electron system considered here, fermionic particle-hole 
fluctuations) that couple to the order-parameter fluctuations and become 
massive in the broken-symmetry phase. Because of the latter property,
the fluctuation contribution to the free energy decreases, which 
ultimately leads to a fluctuation-induced first-order transition. If these
fluctuations are integrated out in some approximation, then a nonanalytic
Landau-like theory can be derived which predicts a discontinuous phase
transition. The modified mean-field theory for the magnetic transition 
mentioned above results from such a procedure. For the superconductor 
and liquid crystal transitions, a similar technique was used 
initially.\cite{Halperin_Lubensky_Ma_74} Later, a renormalization-group
description of fluctuation-induced thermal phase transitions was 
developed.\cite{Chen_Lubensky_Nelson_78} Part of our goal here is to do 
the same for the quantum ferromagnetic transition.

In addition, we perform a renormalization-group analysis of the stability 
of the first-order transition predicted by the generalized mean-field 
theory. It turns out that the first-order transition is stable if it
occurs at a sufficiently large value of the paramter $t$ in 
Eq.\ (\ref{eq:1.2}). However, if it occurs at small values of $t$, then
the first-order transition can in turn become unstable with respect to
fluctuations. The final result in that case is a second-order transition
that is induced by fluctuations, by a mechanism that is similar to the
one discussed in the context of classical Potts models by Fucito and 
Parisi.\cite{Fucito_Parisi_81} This second-order transition is distinct
from Hertz's mean-field transition and belongs to a different universality
class. Depending on microscopic parameter values, the ferromagnetic QPT
in itinerant electron systems can thus be either of first order, or of
second order, in agreement with the experimental observations mentioned
at the beginning of this section. Moreover, the critical behavior in the
continuous case in $d=3$ is mean field-like with logarithmic corrections
to scaling. Within the current experimental accuracy, this is 
indistinguishable from mean-field exponents, again in agreement with the
experimental observations.

The outline of this paper is as follows. In Section \ref{sec:II} we give 
a basic field theory describing coupled magnetization fluctuations and soft 
fermionic degrees of freedom. In Section\ \ref{sec:III} we derive and discuss 
the generalized mean-field theory that results from integrating out the 
fermionic modes and that predicts a first-order transition. In 
Sec.\ \ref{sec:IV} we perform a one-loop renormalization-group analysis of
the field theory. We show that the renormalized quartic coupling constant 
can become negative at large scales, leading to the first-order phase 
transition described by the generalized mean-field theory. However, in a
certain parameter regime fluctuations keep the quartic coefficient
positive, which results in a continuous phase transition.
In Section \ref{sec:V} we further discuss our results, compare them with 
previous work, and comment on the experimental situation. Parts of the results
presented here have been previously announced in two short 
publications.\cite{us_1st_order_PRL,us_1st_order_second_PRL}

\section{The Coupled Field Theory}
\label{sec:II}

Recently we have derived and discussed a local field theory describing the 
quantum ferromagnetic transition in disordered itinerant electron 
systems.\cite{us_fm_local_I,us_fm_local_II} 
This field theory describes the coupling between 
the soft or massless fermionic degrees of freedom (which in a disordered 
electron system are diffusive, i.e., the frequency is a quadratic function 
of the wavenumber) and the magnetization fluctuations. Here we give the 
analogous theory for clean electronic system. In this case the fermionic 
soft modes have a linear dispersion, i.e., the frequency is a linear function 
of the wavenumber. There are numerous ways to construct field theories that 
describe these soft modes; here we choose the method developed in 
Refs.\ \onlinecite{us_fermions_I,us_fermions_II}.

\subsection{Soft Modes}
\label{subsec:II.A}

A generalized Landau-Ginzburg-Wilson (LGW) theory that keeps all of the 
soft modes in the problem will be described in terms of an action 
${\cal A}$ that depends on a field
${\bm M}$ describing the fluctuating magnetization, and on a field $q$
describing the soft fermionic two-particle 
modes.\footnote{Of course the 
magnetization is also if electronic origin. However, since the order 
parameter fluctuations play a special role in the theory, we separate 
them out and refer to all other soft modes as ``fermionic''.}
All other
modes we imagine have been integrated out in order to arrive at the
effective theory. The partition function can then be written in terms
of a functional integral with respect to $\bm{M}$ and $q$,
\begin{subequations}
\label{eqs:2.1} 
\be
Z = \int D[{\bm M},q]\ e^{{\cal A}[{\bm M},q]}\quad.  
\label{eq:2.1a}
\ee
The action will consist of a part that depends only on the magnetization, a 
part that depends only on the fermionic degrees of freedom, and a
coupling between the two, 
\be
{\cal A}[{\bm M},q] = {\cal A}_{M} + {\cal A}_{q} + A_{M,q}\quad.  
\label{eq:2.1b}
\ee
\end{subequations}%
The various pieces of the action in Eq.\ (\ref{eqs:2.1}) can be derived 
starting from a microscopic fermionic action, or more generally written 
down on the basis of symmetry 
arguments.\footnote{Both of these approaches have been widely used in the
history of phase transitions. For instance, the Ginzburg-Landau theory of
superconductivity was originally deduced from general symmetry arguments
and only later derived from the microscopic BCS theory.}
Here we 
choose the latter approach, with occasional references to the miscroscopic 
theory as a check. For a complete derivation from a microscopic action, 
the techniques of Refs.\ \onlinecite{us_fm_local_I} and 
\onlinecite{us_fermions_I} can be used.

${\cal A}_{M}$ is just a static, local, LGW functional for the magnetization 
fluctuations. It is local because no massless modes that couple to the
magnetization have been integrated out,
and it can be chosen static because the relevant (in the long-wavelength, 
low-frequency limit) dynamical part will be shown to be provided by the
coupling to the $q$ fluctuations, see Eq.\ (\ref{eq:2.9c}) below. 
Neglecting terms that are irrelevant for our purposes, ${\cal A}_{M}$ is 
given by
\be
{\cal A}_M = -\int dx\ {\bm M}(x)\,[t - a\nabla^2]\,{\bm M}(x)
             - u\int dx\ {\bm M}^4(x)\quad.  
\label{eq:2.2}
\ee
Here $x\equiv (\bm{x},\tau)$ comprises the real space position ${\bm x}$ 
and the imaginary time $\tau$. $\int dx = \int d{\bm x}\int_0^{\beta} d\tau$ 
with $\beta = 1/k_{\text{B}}T$, where $T$ is the temperature. $t$ is the
dimensionless distance from the bare LGW critical point, and $a$ and $u$
are positive constants. The physical magnetization $m$ is proportional to the 
expectation value of the field ${\bm M}$. 
For later reference we also define a temporal Fourier
transform of the field ${\bm M}$ by
\be
{\bm M}_n({\bm x}) = \sqrt{T}\int_0^{\beta}d\tau\ e^{i\Omega_n\tau}\,
                     {\bm M}(x)\quad,
\label{eq:2.3}
\ee
with $\Omega_n = 2\pi Tn$ a bosonic Matsubara frequency.

The soft fermion field $q$ originates from the composite fermion 
variables\cite{us_fermions_I}
\bse
\label{eqs:2.4}
\be
Q_{12}\cong \frac{i}{2}\left(
\begin{array}{cccc}
-\psi _{1\uparrow }{\bar\psi }_{2\uparrow } & -\psi _{1\uparrow }
{\bar\psi }_{2\downarrow } & -\psi _{1\uparrow }\psi _{2\downarrow } & 
\psi _{1\uparrow }\psi _{2\uparrow } \\ 
-\psi _{1\downarrow }{\bar\psi }_{2\uparrow } & -\psi _{1\downarrow }
{\bar\psi }_{2\downarrow } & -\psi _{1\downarrow }\psi _{2\downarrow }
& \psi _{1\downarrow }\psi _{2\uparrow } \\ 
{\bar\psi }\hspace{0.02in}_{1\downarrow }{\bar\psi }_{2\uparrow }
& {\bar\psi }_{1\downarrow }{\bar\psi }_{2\downarrow } & {\bar
\psi }_{1\downarrow }\psi _{2\downarrow } & - {\bar\psi }_{1\downarrow
}\psi _{2\uparrow } \\ 
-{\bar\psi }_{1\uparrow }{\bar\psi }_{2\uparrow } & - {\bar
\psi }_{1\uparrow }{\bar\psi }_{2\downarrow } & - {\bar\psi }
_{1\uparrow }\psi _{2\downarrow } & {\bar\psi }_{1\uparrow }\psi
_{2\uparrow }
\end{array}\right)\ .
\label{eq:2.4a}
\ee
Here the $\psi$ and ${\bar\psi}$ are the fermionic, i.e., Grassmann-valued, 
fields that provide the basic description of the 
electrons,\cite{Negele_Orland_88}
and all fields are understood to be taken at position $\bm{x}$. The indices
$1$, $2$, etc. denote the dependence of the Grassmann fields on fermionic
Matsubara frequencies $\omega_{n_1} = 2\pi T (n_1+1/2)$, etc., and the arrows
denote the spin projection. It is convenient to
expand the $4\times 4$ matrix in Eq.\ (\ref{eq:2.4a}) in a spin-quaternion 
basis,\cite{Efetov_Larkin_Khmelnitskii_80} 
\be
Q_{12}({\bm x}) = \sum_{r,i=0}^{3}(\tau_r\otimes s_i)\,{_r^iQ}_{12}({\bm x})
                   \quad,  
\label{eq:2.4b}
\ee
with $\tau _0 = s_0 = \openone_2$ the $2\times 2$ unit matrix, and 
$\tau_j = -s_j = -i\sigma_j$ $(j=1,2,3)$, with $\sigma_{1,2,3}$ the Pauli
matrices. In this basis, $i=0$ and $i=1,2,3$ describe the spin-singlet and
spin-triplet degrees of freedom, respectively. The $r=0,3$ components
correspond to the particle-hole channel (i.e., products ${\bar\psi}\psi$
or $\psi{\bar\psi}$), 
while $r=1,2$ describe the particle-particle channel (i.e., products
${\bar\psi}{\bar\psi}$ or $\psi\psi$). For our purposes the
latter are not important, and we therefore drop the $r=1,2$ from the
spin-quaternion basis. In terms of the remaining fields, the spin
density can be expressed as
\bea
n_{\text{s}}^i({\bm x},i\Omega_n) &=& \sqrt{T}\sum_m\sum_{ab}
   {\bar\psi}_{m,a}({\bm x})\,\sigma^i_{ab}\,\psi_{m+n,b}({\bm x})
\nonumber\\
&&\hskip -32pt = \sqrt{T}\sum_m\sum_{r=0,3}(\sqrt{-1})^r \tr\,
   \left[(\tau_r\otimes s_i)\,Q_{m,m+n}({\bm x})\right]\,,
\nonumber\\
&&\hskip 59pt (i=1,2,3)\label{eq:2.4c}
\eea

The matrix elements of $Q$ are bilinear in the fermion fields, so
$Q$-$Q$ correlation functions describe two-fermion excitations.
In a Fermi liquid, the $Q$-fluctuations are massive and soft,
respectively, depending on whether the two frequencies carried by
the $Q$ field have the same sign, or opposite signs, respectively.
We therefore separate the $Q$ fluctuations into massless modes, 
$q_{12}$, and massive modes, $P_{12}$, by
splitting the matrix $Q$ into blocks in frequency space, 
\bea
Q_{nm}({\bm x}) = \Theta(nm)\,P_{nm}({\bm x}) &+& \Theta(n)\Theta(-m)\, 
                                                  q_{nm}({\bm x})
\nonumber\\
          &&\hskip -58pt +\Theta(-n)\Theta(m)\,q^{\dagger}_{nm}({\bm x})\quad.
\label{eq:2.4d}
\eea
\ese%
In what follows, we will incorporate the frequency constraints expressed
by the step functions into the fields $P$ and $q$, respectively. That is,
the frequency indices of $P$ must always have the same sign, and those of
$q$ and $q^{\dagger}$ must always have opposite signs.

Finally, we define spatial Fourier transforms by
\be
M_n({\bm k}) = \frac{1}{\sqrt{V}}\int d{\bm x}\ M_n({\bm x})\quad,
\label{eq:2.5}
\ee
and analogously for the fields $q$ and $q^{\dagger}$.

\subsection{Soft-mode field theory}
\label{subsec:II.B}

The massive modes $P$ can be formally integrated out to obtain an effective
action for the soft modes, $q_{nm}$. This can be done perturbatively,
as the $P$-dependent part of the action takes the form of a stable Gaussian
(i.e., quadratic in $P$) piece, and terms of higher order in $P$ as well
as terms coupling $P$ and $q$, and $P$ and ${\bm M}$, respectively, 
starting with bilinear coupling terms. 
As can be seen from Eq.\ (\ref{eq:2.4a}),
the $q$ are particle-hole excitations, which in a clean electron system have
a linear dispersion relation, i.e., the frequency scales linearly with
the wavenumber. The Gaussian part of the fermionic action will therefore
have the form
\bse
\label{eqs:2.6}
\be
{\cal A}_{q}^{(2)} \hskip -1pt = \hskip -1pt \frac{-4}{G}\!\int\!\hskip -1pt 
       d{\bm x}\,d{\bm y}\!\hskip -1pt\sum_{1,2,3,4} \sum_{r,i}
         {_r^i q}_{12}({\bm x})\,{^i\Gamma}_{12,34}^{(2)}({\bm x}-{\bm y})\,
          {_r^i q}_{34}({\bm y}).
\label{eq:2.6a}
\ee
The vertex function $\Gamma^{(2)}$ is most easily written in momentum space,
\be
{^i\Gamma}_{12,34}^{(2)}({\bm k}) = \delta_{13}\,\delta_{24}\,
    \Gamma^{(2,0)}_{12}({\bm k})
    + \delta_{1-3,2-4}\,\delta_{i0}\,2\pi T G K_{\text{s}}\quad,
\label{eq:2.6b}
\ee
with
\be
\Gamma^{(2,0)}_{12}({\bm k}) = \vert{\bm k}\vert + GH\Omega_{1-2}
   \quad.
\label{eq:2.6c}
\ee
\ese%
Here $G$ and $H$ are model-dependent coefficients.
\footnote{We deliberately use the same notation as in our 
treatment of disordered ferromagnets, Ref.\ \onlinecite{us_fm_local_I}, 
to underscore the similar structures of the two theories.}
If one derives Eq.\ (\ref{eq:2.6b}) from the microscopic model of 
Ref.\ \onlinecite{us_fermions_I}, one finds 
$G = \pi^2 N_{\text{F}}/2v_{\text{F}}$
and $H = 1/\pi N_{\text{F}}$, with $v_{\text{F}}$ the Fermi velocity, and
$N_{\text{F}}$ the density of states per spin at
the Fermi surface. More generally, however, $G$ and $H$ will be arbitrary
coefficients with the appropriate dimensions. $K_{\text{s}}$ is a 
spin-singlet interaction amplitude that we include in our Gaussian theory
in a RPA-type fashion.
Inverting $\Gamma^{(2)}$ shows that its presence does not change the
frequency-momentum structure of the $q$-propagator, see Eqs.\ (\ref{eqs:2.10})
below. There is no spin-triplet interaction in the bare action since its
effects are included in ${\cal A}_M$. In a formal derivation from a
microscopic action, this can be achieved by means of a Hubbard-Stratonovich
decoupling of the spin-triplet interaction, with ${\bm M}$ the
Hubbard-Stratonovich field.\cite{Hertz_76,us_fm_local_I} 
However, as long as $K_{\text{s}}$ is nonzero,
it generates a spin-triplet interaction in perturbation theory. This has
important consequences, see Sec.\ \ref{subsec:III.A} below, and it is
the reason we include $K_{\text{s}}$.

The part of the action coupling ${\bm M}$ and $q$ originates from a term
${\cal A}_{M-Q}$ that couples ${\bm M}$ and $Q$. Such a term must be present 
since in the presence of a magnetization the fermionic spin density will 
couple linearly to it. Using Eq.\ (\ref{eq:2.4c}), we thus obtain
\bse
\label{eqs:2.7} 
\bea
{\cal A}_{M-Q} &=& 2c_1\sqrt{T}\int d{\bm x}\sum_n\sum_{i=1}^3
             M_n^{i}({\bm x})
\nonumber\\
&&\times\sum_{r=0,3}(-1)^{r/2}\sum_{m}\tr[(\tau_{r}\otimes s_{i})\,
   Q_{m,m+n}({\bm x})]\ ,
\nonumber\\  
\label{eq:2.7a}
\eea
with a model-dependent coefficient $c_1$. In a technical derivation from
a model with a point-like spin-triplet interaction amplitude $K_{\text{t}}$,
this term also is produced by a Hubbard-Stratonovich
decoupling of the spin-triplet interaction term,\cite{us_fm_local_I} and 
$c_1 = \sqrt{\pi K_{\text{t}}/2}$. Defining
a symmetrized magnetization field by
\be
b_{12}({\bm x}) = \sum_{i,r}\left(\tau_r\otimes s_i\right)\,
                  {^i_rb}_{12}({\bm x})\quad,
\label{eq:2.7b}
\ee
with components
\bea
{_r^i b}_{12}({\bm x}) &=& (-)^{r/2}\sum_{n}\delta_{n,n_1-n_2}\left[
   M_n^i({\bm x}) \right.
\nonumber\\
&&\hskip 20pt \left. + (-)^{r+1} M_{-n}^i({\bm x})\right]\quad,
\label{eq:2.7c}
\eea
allows to rewrite Eq.\ (\ref{eq:2.7a}) in a more compact form,
\be 
{\cal A}_{M-Q} = c_1\sqrt{T}\int d{\bm x}\ 
                 \tr\left(b({\bm x})\,Q({\bm x})\right)\quad.
\label{eq:2.7d}
\ee
\ese%
Using Eq.\ (\ref{eq:2.4d}) in Eq.\ (\ref{eq:2.7a}) or (\ref{eq:2.7d}),
and integrating out the massive $P$-fluctuations, obviously
leads to a series of terms coupling ${\bm M}$ and $q$, ${\bm M}$
and $q^2$, etc. We thus obtain ${\cal A}_{M,q}$ in form of a series
\bse
\label{eqs:2.8}
\be
{\cal A}_{M,q} = {\cal A}_{M-q} + {\cal A}_{M-q^2} + \ldots
\label{eq:2.8a}
\ee
The first term in this series is obtained by just
replacing $Q$ by $q$ in Eq.\ (\ref{eq:2.7a}),
\bea
A_{M-q} &=& c_1T^{1/2}\int d{\bm x}\ \tr\left(b({\bm x})\,q({\bm x})\right)
\nonumber\\
        &=& 8c_1T^{1/2}\sum_{12}\int d{\bm x}\sum_r\sum_{i=1}^3
            {_r^i b}_{12}({\bm x})\,{_r^i q}_{12}({\bm x})\quad.
\nonumber\\
\label{eq:2.8b}
\eea
The next term in this expansion must have the overall structure 
$$
{\cal A}_{M-q^2} \propto \int d{\bm x}\ \tr \left(b({\bm x})\,q({\bm x})\,
   q^{\dagger}({\bm x})\right)\quad.
$$
The details require information about the structure
of the massive modes that were integrated out in going from $Q$ to $q$.
From the derivation of the nonlinear sigma model that results in the
disordered case if one integrates out the massive $P$-fluctuations in tree 
approximation\cite{Wegner_Schaefer_80,us_fermions_I} 
it is known that the resulting
effective fermion matrix field is traceless, i.e., $(q^2)_{nm}$ in
the above expression enters with different signs depending on whether
$n$ and $m$ are both positive or both negative. This feature carries over
to the clean case and yields
\bea
A_{M-q^2} &=& c_{2}\sqrt{T}\int d{\bm x}\sum_{123}
     \sum_{rst}\sum_{i=1}^{3}\sum_{jk}{_r^i b}_{12}({\bm x)}
\nonumber\\
&&\times\left[{_s^j q}_{23}({\bm x}){_t^k q}_{13}({\bm x})
\tr(\tau_r\tau_s\tau_t^{\dagger})\tr(s_i s_j s_k^{\dagger})\right.
\nonumber\\
&& \left.- {_s^j q}_{32}({\bm x}){_t^k q}_{31}({\bm x})\tr(\tau_r
   \tau_s^{\dagger}\tau_t)\,\tr(s_i s_j^{\dagger} s_k)\right]\ ,
\nonumber\\
\label{eq:2.8c}
\eea
\ese%
with $c_2$ another positive constant. The bare values of $c_1$ and
$c_2$ are related, $c_2 = c_1/16$.
Terms of higher order in $q$
in this expansion will turn out to be irrelevant for determining the
behavior at the quantum phase transition.

\subsection{Gaussian propagators}
\label{subsec:II.C}

We will be interested in the renormalization group flows of the various 
parameters in the field theory defined above. We will need the Gaussian 
propagators of the theory in the paramagnetic phase. These are easily
determined from the quadratic form given by the $M^2$, $q^2$, and $Mq$
parts of the above action. Performing a spatial Fourier transform,
and using the symbol $\langle\ldots\rangle$ for the Gaussian average,
we find for the order parameter correlations
\bse
\label{eqs:2.9} 
\be
\langle M_n^i({\bm k})\,M_m^j({\bm p})\rangle
   = \delta_{{\bm k},-{\bm p}}\,\delta_{n,-m}\,\delta_{ij}\,\frac{1}{2}\,
      {\cal M}_n({\bm k})\quad, 
\label{(2.10a)}
\ee
\bea
\langle{_r^i b}_{12}({\bm k})\,{_s^j b}_{34}({\bm p})\rangle
   &=& -\delta_{{\bm k},-{\bm p}}\,\left[\delta_{1-2,3-4} 
                                  - (-)^r\delta_{1-2,4-3}\right]
\nonumber\\
&&\times \delta_{ij}\,\delta_{rs}\,
         {\cal M}_{1-2}({\bm k})\quad,
\label{eq:2.9b}
\eea
in terms of the paramagnon propagator, 
\be
{\cal M}_n({\bm k}) = \frac{1}{t + a{\bm k}^2 + \frac{(4G c_1^2/\pi)\vert
    \Omega_n\vert}{\vert{\bm k}\vert + GH\vert\Omega_n\vert}}\quad.  
\label{eq:2.9c}
\ee
\ese%
Notice that the coupling between the order parameter field and the fermionic
degrees of freedom has produced the dynamical piece of ${\cal M}$ that is
characteristic of clean itinerant ferromagnets.

For the fermionic propagators we find, 
\bse
\label{eqs:2.10}
\be
\langle{_r^i q}_{12}({\bm k})\,{_s^j q}_{34}({\bm p})\rangle = 
   \delta_{{\bm k},-{\bm p}}\,\delta_{ij}\,\delta_{rs}\,\frac{G}{8}\,
    {^i\Gamma}_{12,34}^{(2)-1}({\bm k)}\quad,  
\label{eq:2.10a}
\ee
with, 
\bea
{^0\Gamma}_{12,34}^{(2)-1}({\bm k}) &=& \delta_{13}\,\delta_{24}\,
   {\cal D}_{1-2}({\bm k})
\nonumber\\
&&\hskip -21pt -\delta_{1-2,3-4}\,2\pi T G K_{\text{s}}{\cal D}_{1-2}
   ({\bm k})\,{\cal D}_{1-2}^{\text{(s)}}({\bm k})\quad,
\nonumber\\
\label{eq:2.10b}
\eea
\bea
{^{1,2,3}\Gamma}_{12,34}^{(2)-1}({\bm k}) &=& \delta_{13}\,\delta_{24}\,
   {\cal D}_{1-2}({\bm k})
\nonumber\\
&&\hskip -50pt -\delta_{1-2,3-4}8T G c_1^2({\cal D}_{1-2}({\bm k}))^2
    {\cal M}_{1-2}({\bm k})\quad,
\nonumber\\
\label{eq:2.10c}
\eea
where ${\cal D}$ and ${\cal D}_{\text{s}}$ are the propagators
\be
{\cal D}_{n}({\bm k}) = \frac{1}{\vert{\bm k}\vert + GH\Omega_n}\quad,
\label{eq:2.10d}
\ee
\be
{\cal D}_{n}^{\text{(s)}}({\bm k}) = \frac{1}{\vert{\bm k}\vert 
   + G(H + K_{\text{s}})\Omega_n}\quad.  
\label{eq:2.10e}
\ee
\ese%
Notice that ${^0\Gamma}^{(2)-1}$ is actually the inverse of ${^0\Gamma}^{(2)}$
given by Eq.\ (\ref{eq:2.6b}), while the analogous statement for 
${^{1,2,3}\Gamma}^{(2)-1}$ is not true. This is because the coupling between
$\bm M$ and $q$ gives an additional contribution to the fermionic spin-triplet
propagator.

Finally, due to the coupling between $\bm M$ and $q$ we also have a mixed 
propagator,
\bea
\langle{_r^i q}_{12}({\bm k})\,{_s^j b}_{34}({\bm p})\rangle\hskip -2pt 
     &=& \hskip -1pt
     -\delta_{{\bm k},-{\bm p}}\left[\delta_{1-2,3-4} 
                        + (-)^{r+1}\delta_{1-2,4-3}\right]
\nonumber\\
&&\hskip -30pt\times\delta_{rs}\,\delta_{ij}\,G\,c_{1}\sqrt{T}
     {\cal D}_{1-2}({\bm k}){\cal M}_{1-2}({\bm k})\ .
\label{eq:2.11}
\eea

\subsection{Higher order terms, and diagram rules for a loop expansion}
\label{subsec:II.D}

The action defined by Eqs.\ (\ref{eqs:2.6}) - (\ref{eqs:2.8})
suffices to extract the information we are interested in, but it is incomplete
from a calculational point of view. Namely, in order to set up a loop
expansion and renormalize the vertices in our action to one-loop order,
one needs the term of order $q^4$. Although it is possible to determine
the desired renormalizations without knowing this term explicitly, see
below, for completeness and later reference we here give such a term that
satisfies basic symmetry requirements.

On general grounds, and by analogy with the disordered 
case,\cite{us_fermions_I} this term must have the structure
\bse
\label{eqs:2.12}
\bea
{\cal A}_q^{(4)}&=&\frac{1}{4G}\int d{\bm x}_1\,d{\bm x}_2\,d{\bm x}_3\, 
    d{\bm x}_4
%   \sum_{1,2\atop 3,4} \sum_{r_1,r_2\atop r_3,r_4} \sum_{i_1,i_2\atop i_3,i_4}
    \sum_{\genfrac{}{}{0pt}{}{1,2}{3,4}}
    \sum_{\genfrac{}{}{0pt}{}{r_1,r_2}{r_3,r_4}}
    \sum_{\genfrac{}{}{0pt}{}{i_1,i_2}{i_3,i_4}}
\nonumber\\
&&\times \tr \left(\tau_{r_1}\tau_{r_2}^{\dagger}\tau_{r_3}
                            \tau_{r_4}^{\dagger}\right)\  
          \tr \left(s_{i_1} s_{i_2}^{\dagger} s_{i_3} s_{i_4}^{\dagger}\right)
\nonumber\\
&&\times \Gamma^{(4)}_{12}({\bm x}_1-{\bm x}_4,{\bm x}_2-{\bm x}_4,{\bm x}_3
                                                                -{\bm x}_4)\,
\nonumber\\
&&\times
   {^{i_1}_{r_1}q}_{12}({\bm x}_1)\,{^{i_2}_{r_2}q}_{32}({\bm x}_2)\,
   {^{i_3}_{r_3}q}_{34}({\bm x}_3)\,{^{i_4}_{r_4}q}_{14}({\bm x}_4)\quad.
\nonumber\\
\label{eq:2.12a}
\eea
The vertex function $\Gamma^{(4)}$ can be expressed in terms of the
two-point vertex $\Gamma^{(2)}$, Eqs.\ (\ref{eq:2.6b},\ref{eq:2.6c}).
\footnote{This structure can be deduced as follows. Integrating
 out the massive modes in tree approximation proceeds formally as in the
 disordered case, where it produces a nonlinear sigma model, see 
 Refs.\ \onlinecite{Wegner_Schaefer_80,us_fermions_I}. The only difference
 is that in the clean case, soft single-particle excitations have been
 integrated out to produce the effective $Q$-field theory. This leads to
 a singular vertex in the gradient-squared term of the sigma model that
 is, in the long-wavelength limit, proportional to an inverse wavenumber
 and changes the dispersion relation of the soft modes from a diffusive
 one to a linear one, see Eq.\ (\ref{eq:2.6c}). The requirement that
 the two-point vertex $\Gamma^{(2)}$ remains soft under renormalization
 puts constraints on the higher vertices in the $q$-expansion. These
 constraints are fulfilled if the singular vertex is the same for
 all terms in the $q$-expansion. This suggests that the $q^4$ vertex, 
 if expressed in terms of the $q^2$ vertex, is the same as in the nonlinear 
 sigma model. This procedure leads to Eqs.\ (\ref{eqs:2.12}). In 
 Sec.\ \ref{sec:IV}
 we will see that the results obtained from this $\Gamma^{(4)}$ agree with
 those obtained by other, more indirect, means. We also mention that the
 above arguments suggest that one can construct an effective field theory
 for the soft modes in a clean fermion system that is analogous, and closely
 related, to the well-known nonlinear sigma model that describes disordered
 fermions. This general theory will be pursued separately.}
In Fourier space, and neglecting $K_{\text{s}}$, it reads
\be
\Gamma^{(4)}_{12}({\bm k}_1,{\bm k}_2,{\bm k}_3) = \frac{1}{2}\,\left[
   \Gamma^{(2,0)}_{12}({\bm k}_1 + {\bm k}_2)
   + \Gamma^{(2,0)}_{12}({\bm k}_2 + {\bm k}_3) \right].
\label{eq:2.12b}
\ee
\ese%
In addition there are terms of order $q^3$ and $q^4$ that are proportional
to $K_{\text{s}}$, as well as terms of higher order in $q$, but they will
not be important for our purposes.

As the last step in defining our effective field theory, we need to
remember that setting up a $q$-field theory requires a Lagrange multiplier
field $\lambda$ that constrains bilinear products of the underlying fermion
fields to the classical matrix field $Q$. In clean systems, the $\lambda$-field
is soft with a propagator that is given by minus the noninteracting part of
the $q$-propagator,\cite{us_fermions_I}
\be
\langle{_r^i\lambda}_{12}({\bm k})\,{_s^j\lambda}_{34}({\bm p})\rangle =
   - \delta_{{\bm k},-{\bm p}}\,\delta_{ij}\,\delta_{rs}\,
   \delta_{13}\,\delta_{24}\,\frac{G}{8}\,{\cal D}_{1-2}({\bm k})\quad.
\label{eq:2.13}
\ee
This field couples to $q$ in a way that results, upon integrating out
$\lambda$, in the following diagram rules\cite{us_fermions_I}

\smallskip
{\it Rule 1.} For calculating propagators in a loop expansion, all
internal $q$-propagators must be taken as the interacting part of the
Gaussian propagator, i.e., as the second term on the right-hand side of
Eq.\ (\ref{eq:2.10b}) or (\ref{eq:2.10c}). 

\smallskip
{\it Rule 2.} For calculating vertex functions, Rule 1 also applies.
In addition, one needs to consider all reducible diagrams (which normally
do not contribute to the vertices), with all reducible propagators replaced
by the $\lambda$-propagator, Eq.\ (\ref{eq:2.13}).

\begin{figure}
\includegraphics[width=6cm]{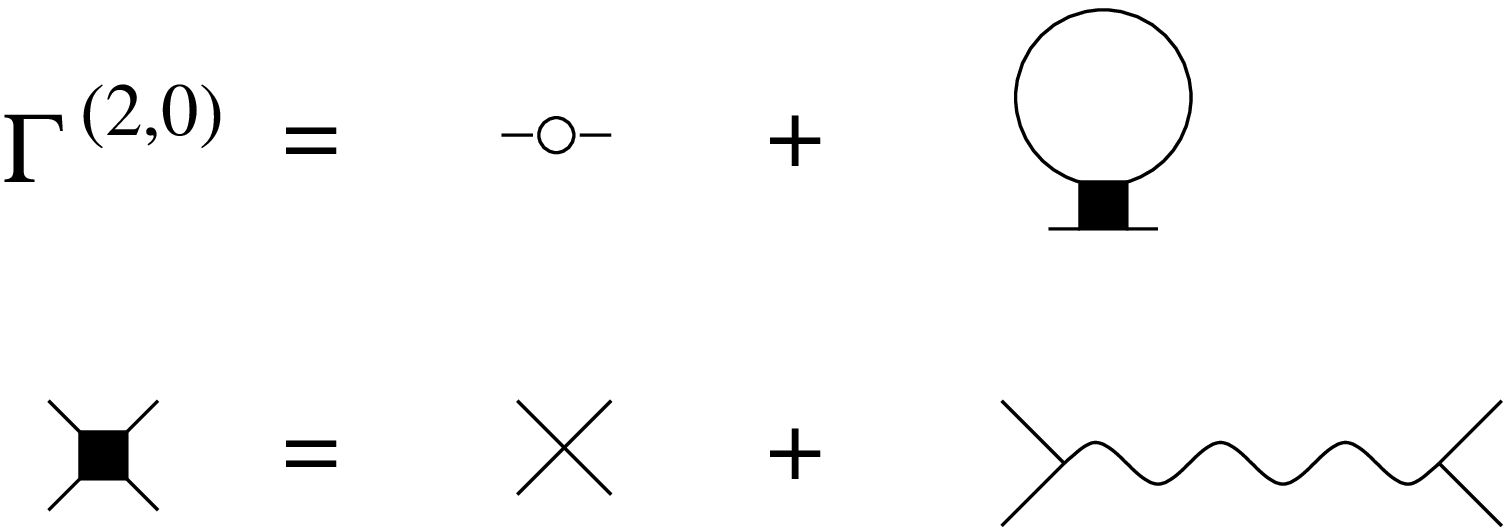}
\caption{\label{fig:1} The noninteracting part of the two-point $q$-vertex 
 to one-loop order. Solid lines denote the interacting part of the 
 $q$-propagator, and the wave line denotes the $\lambda$-propagator. See the 
 text for further explanation.}
\end{figure}

\smallskip
As an illustration, we show in Fig.\ \ref{fig:1} the diagrams for the 
renormalization of $\Gamma^{(2,0)}$, Eq.\ (\ref{eq:2.6c}), to one-loop order.

This completes the definition of our effective action, and we will now
proceed to discuss the ferromagnetic transition it describes.

\section{Generalized Mean-Field Theory, and the First-Order Phase Transition}
\label{sec:III}

In this section we derive a generalized mean-field theory for the 
ferromagnetic transition in low-temperature itinerant electron systems. 
It structurally maps onto the generalized mean-field theory for the 
superconducting transition at finite temperature.\cite{Halperin_Lubensky_Ma_74} 
The transition predicted by
these theories is of first order. We then discuss the conditions under which
this result is stable. We will see that, contrary to the usual concepts
concerning first-order phase transitions, the mean-field description can
be invalidated by fluctuation effects that drive the transition second order.
Physically, the first-order transition turns out to be unstable when it is
too close to a second-order transition with sufficiently strong fluctuations,
otherwise it is stable.

\subsection{Generalized mean-field theory}
\label{subsec:III.A}

An effective action, $A_{\text{eff}}[{\bm M}]$, involving only the magnetization 
order parameter, can be obtained by integrating out the fermion fields,
\be
e^{{\cal A}_{\text{eff}}[{\bm M}]} = \int D[q]\ e^{{\cal A}[{\bm M},q]}\quad.
\label{eq:3.1}
\ee
Here ${\cal A}$ is the action given by Eq.\ (\ref{eq:2.1b}).
In general the evaluation of this expression is very difficult. However,
it can be evaluated exactly within a generalized mean-field approximation,
which is defined as follows. First, 
we ignore temporal and spatial variations of ${\bm M}$,
\be
M_n^i({\bm x}) \approx \delta_{i3}\,\delta_{n0}\,m/\sqrt{T}\quad.
\label{eq:3.2}
\ee
Second, we restrict ourselves to Gaussian or quadratic order in $q$. 
That is, we replace the full action ${\cal A}$ by 
Eqs.\ (\ref{eq:2.2},\ref{eqs:2.6}, \ref{eqs:2.8}), 
and in these expressions we replace ${\bm M}$ by Eq.\ (\ref{eq:3.2}).

With the bare Gaussian action as given in Sec.\ \ref{subsec:II.B}, and
taking into account the diagram rules mentioned at the end of
Sec.\ \ref{subsec:II.D}, there is no coupling between the soft modes
and the order parameter. However, one needs to acknowledge that under
renormalization, the action ${\cal A}_q$ will acquire a spin-triplet
interaction that is generated as long as $K_{\text{s}}\neq 0$. Let the such
generated interaction constant be ${\tilde K}_{\text{t}}$. Then the 
fermionic 2-point vertex function, Eq.\ (\ref{eq:2.6b}), gets generalized
to
\begin{eqnarray*}
{^i\Gamma}_{12,34}^{(2)}({\bm k})&=&\delta_{13}\,\delta_{24}\,
    (\vert{\bm k}\vert + GH\vert\Omega_{1-2}\vert)
\nonumber\\
&& \hskip 0pt +\ \delta_{1-3,2-4}\,\delta_{i0}\,2\pi T G K_{\text{s}}
\nonumber\\
\end{eqnarray*}
\vskip -40pt
\be
\hskip 94pt +\ \delta_{1-3,2-4}\,(1-\delta_{i0})\,2\pi T G 
     {\tilde K}_{\text{t}}\quad.
\tag{2.6b'}
\label{eq:2.6b'}
\ee

This renormalization-generated spin-triplet interaction leads to a coupling
between the soft fermionic modes and the order parameter in the free energy.
In the resulting generalized mean-field approximation
one obtains for the free energy density, $f(m) = -T{\cal A}_{\text{eff}}/V$,
\bse
\label{eqs:3.3}
\bea
f &=& f[m=0] + t\,m^2 + u\,m^4
\nonumber\\
&&      + \frac{2}{V}\sum_{{\bm k} < \Lambda}
      T\sum_{n}\ln N({\bm k},\Omega_n;m)\quad,
\label{eq:3.3a}
\eea
where $\Lambda$ is an ultraviolet momentum cutoff, and
\bea
N({\bm k},\Omega_n;m) &=& 16 c_2^2 G^4{\tilde K}_{\text{t}}^2 m^2 \Omega_n^2
\nonumber\\
&&\hskip -1pt + (\vert{\bm k}\vert + GH\Omega_n)^2
     (\vert{\bm k}\vert + G(H + {\tilde K}_{\text{t}})\Omega_n)^2
\nonumber\\
\label{eq:3.3b}
\eea
Minimizing $f$ with respect to the magnetization gives the equation of state,
\bea
h &=& 2\,t\,m + 4\,u\,m^3 
\nonumber\\
&& +\, m\,64\,c_2^2\,G^4 {\tilde K}_{\text{t}}^2 
      \frac{1}{V}\sum_{{\bm k} < \Lambda} T\sum_{n=1}^{\infty}
   \frac{\Omega_n^2}{N({\bm k},\Omega_n;m)}\quad,
\nonumber\\
\label{eq:3.3c}
\eea
\ese%
with $h$ an external magnetic field.

\subsection{Discussion of the generalized mean-field equation of state}
\label{subsec:III.B}

We start with some general comments regarding the result, 
Eqs.\ (\ref{eqs:3.3}).
The last term in both Eqs.\ (\ref{eq:3.3a}) and (\ref{eq:3.3c}) arises from 
fermionic fluctuations, namely, the
$^i_rq$ with $r=0,3$ and $i=1,2$, that are massless in the paramagnetic 
phase, but that become massive in the ordered phase. 
\footnote{This can be seen explicitly by using Eq.\ (\ref{eq:3.2}) in
 Eq.\ (\ref{eq:2.8c}) and re-calculating the $q$-propagator. Notice that
 the mean magnetization $m$ acts like an external magnetic field, which
 breaks the symmetry in spin space and gives two of the three soft 
 spin-triplet modes a mass. This is the clean analog of the
 ``spin diffusons'' in the disordered case, which also acquire a mass in
 either an external magnetic field, or in a phase with a nonvanishing
 magnetization, see, e.g., Refs.\ \onlinecite{us_R,us_fm_mit_I}.}
As discussed elsewhere,\cite{us_chi_s} these fluctuations lead to long-range 
correlations in paramagnetic metals, and to nonanalyticities in either 
the temperature or the wavenumber depence of correlation functions, 
for example, the magnetic susceptibility. It is also interesting to note
that Eqs.\ (\ref{eqs:3.3})
are identical to the equations describing the first-order phase transition
in conventional superconductors at finite 
temperature.\cite{Halperin_Lubensky_Ma_74} As mentioned in the Introduction, 
the physics of the respective phase transitions is very similar as well.

\begin{figure}
\includegraphics[width=5cm]{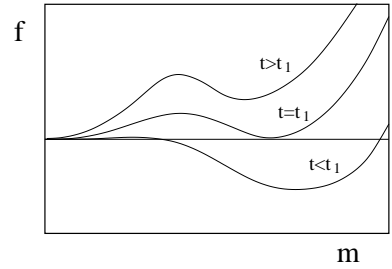}
\caption{\label{fig:2} Schematic form of the free energy as a function of 
 the order parameter.}
\end{figure}

With some work, the integrals or sums in Eqs.\ (\ref{eqs:3.3}) can be 
explicitly performed. However, the most important features can be obtained
by inspection and simple asymptotic analysis. At zero temperature,
and for small $m$, the leading nonanalytic $m$-dependence is a negative 
term on the right-hand side of  Eq.\ (\ref{eq:3.3c}) that is of order $m^d$ 
in generic dimensions, and of order $m^3 \ln 1/m$ in
$d=3$. At low but finite temperatures this nonanalyticity is effectively 
replaced by a negative term of order 
$$m(m^2 + {\text{const.}}\times T^2)^{(d-1)/2}$$ 
in generic dimensions, or 
$$m^{3}\ln 1/(m^2 + {\text{const.}}\times T^2)^{1/2}$$ 
in $d=3$.
\footnote{Notice that there is no term proportional to
$m\,T^2\ln T$ in $d=3$. In other words, there is no renormalization of
$t$, or the magnetic susceptibility, that is proportional to $T^2\ln T$.
This feature of the generalized mean-field theory is in agreement with
exact perturbative calculations, Ref.\ \onlinecite{us_chi_s}, as well
as with Landau Fermi-liquid theory, Ref.\ \onlinecite{Carneiro_Pethick_77}.}
Here const. is a positive constant proportional to
$H^2 (1 + H/{\tilde K}_{\text{t}})^2/c_2^2$. Analogous terms, with an 
extra factor of $m$, 
appear in Eq.\ (\ref{eq:3.3a}). As $T\rightarrow 0$ Eq.\ (\ref{eq:3.3a}) thus
has the standard form of a free energy, or effective potential, 
that leads to a discontinuous phase transition at some $t=t_1>0$, 
see Fig. \ref{fig:2}.
Schematically, this 
free energy functional in the presence of an external magnetic field $h$ 
can be written, in $1<d<3$,
\bse
\label{eqs:3.4}
\bea
f &=& f(m=0) + t\,m^2 - v\,m^2 (m^2 + T^2)^{(d-1)/2}
\nonumber\\
&& + u\,m^4 - hm + \ldots
\label{eq:3.4a}
\eea
and in $d=3$,
\bea
f &=& f(m=0) + t\,m^2 + v\,m^4\ln (m^2+T^2) 
\nonumber\\
&& + u\,m^4 - hm + \ldots
\label{eq:3.4b}
\eea
In this schematic representation, the mean-field equation of state in 
the most interesting case, $d=3$, takes the form
\bea
h &=& 2\,t\,m + 4\,v\,m^3\ln (m^2 + T^2) 
\nonumber\\
&& +\, m^3\,\left(4u + 2v\,\frac{m^2}{m^2 + T^2}\right) \quad.
\nonumber\\
\label{eq:3.4c}
\eea
\ese%
In these equations we use units such that $f$, $m$, and $T$ are measured
in terms of a microscopic energy, e.g., the Fermi energy. $t$, $v$, and
$u$ are then all dimensionless. $v>0$ is quadratic in $K_{\text{t}}$ or 
$c_1^2$, so in
strongly correlated systems $v$ is larger than in weakly correlated ones.

\begin{figure}
\includegraphics[width=5.0cm]{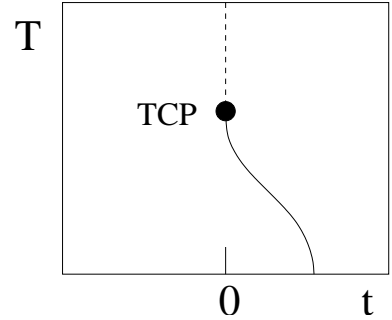}
\caption{\label{fig:3} Schematic form of the phase diagram at $h=0$. 
 The dashed line denotes a second-order transition, the solid line 
 denotes a first-order transition, and TCP denotes the tricritical 
 point.}
\end{figure}

In $d=3$, these equations predict the phase diagram shown in 
Fig.\ \ref{fig:3}.
There is a tricritical point at
\bse
\label{eqs:3.5}
\be
T = T_{\text{tc}} = \exp (-u/2v)\quad.
\label{eq:3.5a}
\ee
At $T=0$, there is a first-order phase transition at $t=t_1$, with the
magnetization changing discontinuously from zero to a value $m_1$. One
finds
\bea
m_1 &=& \exp [-\frac{1}{2}\,(1+u/v)]\quad,
\label{eq:3.5b}\\
t_1 &=& v\,m_1^2\quad.
\label{eq:3.5c}
\eea
\ese%

In $d=2$, there is no finite-temperature magnetic phase transition. However,
at zero temperature there is a QPT, which is predicted by the 
Eqs.\ (\ref{eqs:3.4}) to
be discontinuous. The discontinuity in the magnetization and the transition
point are given by 
\bse
\label{eqs:3.6}
\bea
m_1 &=& (3v/4u)^2\quad,
\label{eq:3.6a}\\
t_1 &=& \frac{u}{3}\,m_1^2\quad.
\label{eq:3.6b}
\eea
\ese%

In $d>3$ the nonanalyticitic terms produced by the soft modes are subleading,
and the transition is described by ordinary mean-field theory. The generalized
mean-field theory thus suggests an upper critical dimension 
$d_{\text{c}}^{\,+}=3$. As we will see in the next section, a more 
sophisticated analysis confirms this result.

\subsection{Validity of the mean-field description}
\label{subsec:III.C}

\begin{figure}
\includegraphics[width=4cm]{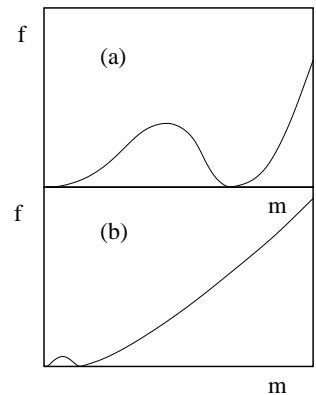}
\caption{\label{fig:4} Schematic forms of the free energy as a function of
 the order parameter. In (a) the first-order transition is not affected by
 fluctuations, in (b) it may be.}
\end{figure}

Normally, first-order phase transitions are not sensitive to fluctuation
effects. We now argue, however, that in the present case of a first order
transition driven by fluctuations (viz., soft fermion modes), additional 
fluctuation effects can destabilize the mechanism underlying the first-order 
transition and lead to a fluctuation-driven second-order transition. This will
happen if, in a sense described below, the first-order transition is too close
to an unrealized second-order one. To illustrate
this point, consider the two schematic free energy functionals shown in
Fig.\ \ref{fig:4}. In Fig.\ \ref{fig:4}(a) the first-order transition
occurs far from $t=0$, and fluctuation effects are negligible. However,
for the case shown in Fig.\ \ref{fig:4}(b) the fluctuations near the
(unrealized) second-order transition at $t=0$ can affect the first-order
transition that preempts the second-order one, and need to be taken into
account.

Before presenting technical details in the next section, let us elaborate
on this general point.
First, we note that as long as one is far from any continuous transition
(which brings in new fluctuation effects) the functional forms of the free 
energy funtions given by Eqs.\ (\ref{eqs:3.4}) are exact for small 
magnetizations. This follows from the properties of a Fermi-liquid fixed point
and the corrections to scaling near it.\cite{us_fermions_I} The mean-field 
description above suggest a second-order, or continuous, phase transition 
at $t=0$, which is preempted by the first-order transition. The latter results
from fluctuations that are germane to a Fermi liquid and have nothing to do
with any critical point. These observations imply that the first-order 
transition discussed above will take necessarily take place if (1) $m_1$ is 
small enough, and (2) $t_1$ is not too small, so that additional fluctuation 
effects due to the underlying critical point at $t=0$ can be ignored. Examining 
the Eqs.\ (\ref{eqs:3.4}) and (\ref{eqs:3.5}) we see that this can occur when 
both $v$ (which describes correlation effects) and $u/v$ are large.
More generally, it is reasonable to expect a first-order phase transition, 
with no restrictions on $m$, whenever correlation effects are large.

Second, the next natural question is, what happens if this is not the case?
In particular, we note the following. The Eqs.\ (\ref{eqs:3.3}) imply that the 
coefficient $v$ is inversely proportional to $H$. As explained in 
Section\ \ref{sec:IV} below, $H$ is proportional to the specific heat 
coefficient. Since $H$ is expected to be sensitive to critical fluctuations,
and perturbation theory suggests a logarithmic divergence at a continuous
transition in $d=3$, this suggests that $v$ might be suppressed close to, 
and at, a continuous transition. To examine this possibility one needs to go 
beyond simple perturbation theory. In the next Section renormalization group
methods are used to untangle the possibilities. We find that a continuous
transition does indeed occur if $t_1$ is sufficiently small, and if the bare 
$u$ is sufficiently large to stabilize the effects of critical fermionic 
fluctuations that are otherwise suppressed.

\section{Renormalization Group Flows, and the second-order Phase Transition}
\label{sec:IV}

\subsection{Renormalization of the effective action}
\label{sec:IV.A}

The parameters $t$, $a$, $u$, $G$, $H$, $c_{1}$, $c_{2}$ as well as the fields
${\bm M}$ and $q$ in the theory defined above are renormalized under 
renormalization group
(RG) transformations. We will employ a differential momentum-shell RG and
integrate over all frequencies. If $b$ is the RG length rescaling factor,
then we rescale wavenumbers and the two fields via
\bse
\label{eqs:4.1}
\bea
{\bm k}&\rightarrow&b{\bm k}\quad,
\label{eq:4.1a}\\
{\bm M}_n({\bm x})&\rightarrow& b^{(d-2+\eta_M)/2}{\bm M}_n({\bm x})\quad,
\label{eq:4.1b}\\
q_{nm}({\bm x})&\rightarrow& b^{(d-2+\eta_q)/2}q_{nm}({\bm x})\quad.  
\label{eq:4.1c}
\eea
Here $\eta_q$ and $\eta_M$ are exponents that characterize the spatial
correlations of the fermion and order-parameter fields, respectively.
The rescaling of imaginary time, frequency, or temperature is less
straightforward. We need to acknowledge the fact that there are two
different time scales in the problem, namely, one that is associated
with the critical order-parameter fluctuations, and one that is associated
with the soft fermionic fluctuations. Accordingly, we must allow for two
different dynamical exponents, $z_M$ and $z_q$, and imaginary time and
temperature may get rescaled either via
\be
\tau \rightarrow b^{-z_M}\tau\quad,\quad T \rightarrow b^{z_M} T\quad,
\label{eq:4.1d}
\ee
or via
\be
\tau \rightarrow b^{-z_q}\tau\quad,\quad T \rightarrow b^{z_q} T\quad,
\label{eq:4.1e}
\ee
\ese%
How these various exponents should be chosen is discussed 
below.
\footnote{We use Ma's method for identifying simple RG fixed points.
 Accordingly, we use physical arguments to choose the values of various
 exponents, and then check self-consistently whether these choices indeed lead
 to appropriate fixed points. See Ref.\ \onlinecite{Ma_76}.}

\subsubsection{Zero-loop flows}
\label{subsubsec:IV.A.1}

In the tree, or zero-loop, approximation the RG flow equations for
the parameters in our field theory are easily determined by power counting
from the action given in Sec.\ \ref{subsec:II.B}. With $\ell = \ln b$ we find
\bse
\label{eqs:4.2}
\bea
\frac{dt}{d\ell} &=& (2-\eta_M)\,t\quad,  
\label{eq:4.2a}\\
\frac{da}{d\ell} &=& -\eta_M\, a\quad,  
\label{eq:4.2b}\\
\frac{du}{d\ell} &=& -(d + z_M + 2\eta_M - 4)\,u\quad,  
\label{eq:4.2c}\\
\frac{d G}{d\ell} &=& -(1 - \eta_q)\,G\quad,  
\label{eq:4.2d}\\
\frac{d H}{d\ell} &=& (2 - z_q - \eta_q)\,H\quad,  
\label{eq:4.2e}\\
\frac{d c_1}{d\ell} &=& \frac{1}{2}\,(4 - z - \eta_q - \eta_M)\,c_1\quad,
\label{eq:4.2f}\\
\frac{d c_2}{d\ell} &=& \frac{1}{2}\,(6 - d - z - 2\eta_q - \eta_M)\,c_2\quad.  
\label{eq:4.2g}
\eea
\ese%

Now we examine these flow equations in order to see whether the allow for
a critical (i.e., unstable in only one direction) fixed point, at least 
above some upper critical dimension. This will amount to an analysis of
the stability, or otherwise, of Hertz's critical fixed point.\cite{Hertz_76}
Note that in giving Eqs.\ (\ref{eq:4.2f}) and (\ref{eq:4.2g}), the 
particular choice of $z$ was not yet specified because it is not obvious if a 
$z_q$ or a $z_M$ should be used for these terms that describe a coupling 
between $q$ and $M$ fields. For the analogous disordered electron problem, 
this point has been discussed in detail in Ref.\ \onlinecite{us_fm_local_I}. 

If we assume the Fermi-liquid degrees of freedom to be at a stable 
Fermi-liquid fixed point, we must choose $G$ and $H$ to be marginal, 
which implies 
\be
\eta_q = 1\quad,\quad z_q = 1\quad.  
\label{eq:4.3}
\ee
Hertz\cite{Hertz_76} further chose (in our language) $a$ and $c_1$ to be 
marginal, which implies
\bse
\label{eqs:4.4}
\be
\eta_M = 0\quad,  
\label{eq:4.4a}
\ee
and 
\be
z_M = 3\quad.
\label{eq:4.4b}
\ee
\ese%
The latter choice is motivated by the paramagnon propagator, 
Eq.\ (\ref{eq:2.9c}), which yields appropriate mean-field critical behavior
only for a marginal $c_1$, given that $G$ and $H$ are marginal. This
also implies that $z = z_M$ in Eq.\ (\ref{eq:4.2f}).

With these choices, $t$ is the relevant variable characterizing the
critical fixed point, and Eq.\ (\ref{eq:4.2a}) yields a correlation 
length exponent $\nu=1/2$. The variable $u$ is irrelevant for $d>1$, 
suggesting an upper critical dimension equal to unity. Indeed,
Hertz's conclusion was that the mean-field fixed point characterized
by the above exponent was stable for $d>1$.

\begin{figure}
\includegraphics[width=4cm]{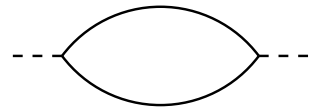}
\caption{\label{fig:5} Example of a fermionic loop renormalizing the
 paramagnon propagator, with the dashed line representing the magnetization
 field. Since the loop integral is over fermionic propagators only, the $c_2$ 
 associated with the vertices carry a time scale given by $z_q$.}
\end{figure}

However, we still need to examine the behavior of $c_2$. It is irrelevant 
for $d>1$ if we use $z = z_M$ in Eq.\ (\ref{eq:4.2g}). However, as already 
emphasized in Ref.\ \onlinecite{us_fm_local_I}, one
also has to consider the case $z=z_q$ in this equation. This becomes obvious
if one uses the $M$-$q^2$ vertex, whose coupling constant is $c_2$, to
construct loops. Clearly, pure fermion loops appear, the simplest example
of which is shown in Fig.\ \ref{fig:5}, and in this case $z=z_q$
is the appropriate choice. 
We illustrate this point below by means of an
explicit calculation. 

Using $z=z_q$ in Eq.\ (\ref{eq:4.2g}), we see 
that $c_2$ becomes relevant with respect to Hertz's fixed point for $d<3$, 
giving an actual upper critical dimension $d_c^{\, +} = 3$. 
This is in agreement with the result of the
generalized mean-field theory, see Sec.\ \ref{subsec:III.B}.
Physically, this surprising results means that soft
or gapless fermion excitations play an important role in determining the phase
transition behavior at, and below, three spatial dimensions even though
naive power counting suggests $d_c^{\,+} = 1$. This is further examined in the 
next subsection, as well as in Section\ \ref{subsec:IV.B}.

\subsubsection{One-loop flows}
\label{subsubsec:IV.A.2}

In this subsection we examine the effects of $c_2$ on the phase transition.
We will be mainly concerned with the behavior in three-dimensions,
the behavior in $d<3$ will be discussed using other techniques in 
Sec.\ \ref{subsec:IV.B} below.

In $d=3$ the relevant diagrams can, in principle, give logarithmic
corrections or renormalizations to the various coupling constants. Taking
into account that there are two time scales, it is easy to show by power
counting that there will be no logarithmic corrections to $c_1$, $c_2$, 
$G$, and $t$. This implies that for these coupling constants, the
flow equations given in Eqs.\ (\ref{eqs:4.2}) remain valid to one-loop order. 
Motivated by the disordered case, we
will be looking for a fixed point where $G$ is marginal, which implies
\bse
\label{eqs:4.5}
\be
\eta_q = 1\quad. 
\label{eq:4.5a}
\ee
We further require $c_1$ (with $z=z_M$) and $c_2$ (with $z=z_q$) to be 
marginal, which implies 
\be
z_M + \eta_M = 3\quad,
\label{eq:4.5b}
\ee
and
\be
z_q + \eta_M = 1\quad.
\label{eq:4.5c}
\ee
\ese%
Of the various scale dimensions introduced above, this leaves only one,
e.g. $\eta_M$, as independent. For the irrelevant version of $c_2$
(with $z = z_M$), Eqs.\ (\ref{eq:4.5a}) and (\ref{eq:4.5b}) imply
\be
\frac{d c_2}{d\ell} = -c_2\quad.
\label{eq:4.6}
\ee

\begin{figure}
\includegraphics[width=4cm]{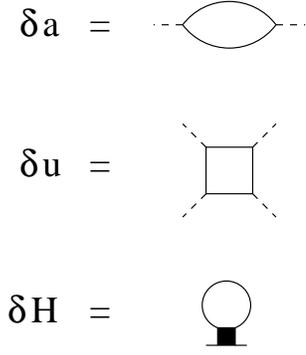}
\caption{\label{fig:6} Diagrams that give the leading renormalization of
 $a$, $u$, and $H$, to one-loop order in $d=3$. The quartic vertex in
 the diagram for $\delta H$ was defined in Fig.\ \ref{fig:1}.}
\end{figure}

For the remaining quantities, power counting shows that they do allow
for logarithmic renormalizations in $d=3$. The diagrams that give rise
to these are shown in Fig.\ \ref{fig:6}.
A crucial feature is that $u$ is
renormalized by a {\em negative} logarithmic term. In a purely perturbative
treatment, this implies that $u$ changes sign, which in turn implies a 
fluctuation driven first-order phase transition, and the existence of a 
tricritical point at finite temperatures, consistent with the generalized
mean-field theory. However, the renormalization group flow equations resum
perturbation theory in a specific way, and in this subsection we show that
this `tricritical' behavior does not necessarily persist to all orders.

The explicit flow equations are obtained by evaluating the diagrams
shown in Fig.\ \ref{fig:6}. Determining the general structure of the flow 
equations does not require a detailed calculation, but can be achieved by
power counting. At zero temperature, we find
\bse
\label{eqs:4.7}
\bea
\frac{d a}{d\ell} &=& -\eta_M a - \frac{A_a}{H}\quad, 
\label{eq:4.7a}\\
\frac{d u}{d\ell} &=& -(2+\eta_M)u - A_u\frac{c_2^2}{H}\quad,  
\label{eq:4.7b}\\
\frac{d H}{d\ell} &=& \eta_M H + \frac{A_H}{a+t}\quad,  
\label{eq:4.7c}
\eea
\ese%
where the $A_i$ are positive constants. 
In giving Eqs.\ (\ref{eqs:4.7}) we have absorbed the marginal coupling
constant $G$ and the marginal version of $c_2$ into these constants.
\footnote{In the case of Eq.\ (\ref{eq:4.7c}), this requires some 
 explanation,
 as the $c_2$ that appears here is nominally the irrelevant $c_2$ (with
 $z = z_M$). However, it effectively acts like a marginal operator since
 the vertex function $\Gamma^{(2)}$ is proportional to a frequency rather
 than being a constant. This mechanism for a nominally irrelevant operator
 turning into a marginal one is the same as in the disordered case and has
 been discussed in detail in Ref.\ \onlinecite{us_fm_local_I}.}

The prefactors $A_i$ can be determined by a detailed calculation of the
diagrams. In the case of $A_a$ and $A_H$, one can also obtain the
result by the following alternative method. $a$ is the coefficient of
the gradient squared term in the spin susceptibility of a non-magnetic
reference system. The logarithmic renormalization of the latter in $d=3$
has been calculated in Ref.\ \onlinecite{us_chi_s}, and we can thus
find the renormalization of $a$ from that paper. For $H$, we notice that
it is related to the specific heat coefficient $\gamma_V = C_V/T$ by
\be
\gamma_V = 8\pi H/3\quad.
\label{eq:4.8}
\ee
This relation between the frequency coupling constant and the 
specific heat was first established for disordered electron 
systems by Castellani and Di Castro.\cite{Castellani_DiCastro_86} A proof
by means of Ward identities\cite{Castellani_et_al_88} applies to clean
systems as well. One can therefore obtain the renormalization of $H$
from a calculation of the specific heat, which in turn follows from
the Gaussian free energy density $f_{\text{G}}$. From Sec.\ \ref{subsec:II.C}
we find, at criticality,
\be
f_{\text{G}} = \frac{3T}{2V}\sum_{\bm p}\sum_{i\Omega_n}
 \ln\left( a{\bm p}^2 + 4Gc_1^2/\pi\vert{\bm p}\vert\right)\quad,
\label{eq:4.9}
\ee
and the specific heat coefficient is obtained by differentiating twice
with respect to temperature.

We have chosen the second method to calculate $A_a$ and $A_H$, and obtain
\footnote{Here we use the fact that the marginal version of $c_2$ is related
to the marginal operator $c_1$, see the remark after Eq.\ (\ref{eq:2.8c}).}
\bse
\label{eqs:4.10}
\bea
A_a = Gc_2^2/9\pi^3\quad,
\label{eq:4.10a}\\
A_H = 3Gc_2^2/\pi^3\quad.
\label{eq:4.10b}
\eea
\ese%
For later reference we note that $A_H/A_a = 27 > 1$, 
but for now we consider the
general case. The value of $A_u$ will not be needed, other than that it is
positive.

Next we solve the Eqs.\ (\ref{eqs:4.7}), first for the case $t=0$. 
It is convenient to first construct flow equations for the quantities
\bse
\label{eqs:4.11}
\be
f = c_2^2/H\quad,\quad g = aH\quad.  
\label{eq:4.11a}
\ee
The flow equations for these objects in $d=3$ are,
\bea
\frac{d g}{d\ell} &=& A_H - A_a\quad,  
\label{eq:4.11b}\\
\frac{d f}{d\ell} &=& -(2+\eta_M)f - A_H f/g\quad.  
\label{eq:4.11c}
\eea
\ese%
Solving these equations and using the result in Eq.\ (\ref{eq:4.7b}) gives
\bse
\label{eqs:4.12}
\be
u(\ell) = e^{-\kappa\ell}\left\{{u_0} - \frac{A_u f_0}{A(k-1)}
          \left[1 - \frac{1}{(1+A\ell)^{K-1}}\right]\right\}\quad,  
\label{eq:4.12a}
\ee
with
\be
\kappa = 2 + \eta_M,\quad A = \frac{A_H - A_a}{g_0}\quad,\quad
     K = \frac{A_H}{A_H - A_a}\quad,
\label{eq:4.12b}
\ee
and
\bea
g_0 = g(\ell=0) = 1/96\pi v_{\text{F}}\quad,
\label{eq:4.12c}\\
f_0 = f(\ell=0) = \pi/16\quad.
\label{eq:4.12d}
\eea
\ese%

Since $K\geq 1$ and $A>0$ for $A_H > A_a$, we see that in contrast to the
perturbative result, $u(\ell)$ does not necessarily become negative as
$\ell\rightarrow\infty$. Rather, the term in braces in Eq.\ (\ref{eq:4.12a})
asymptotically approaches a value $u_0 - A_u f_0 g_0/A_a$. Depending on
microscopic parameter values, $u$ thus may or may not become negative
for large scales. We conclude that a nontrivial continuous phase
transition may exist for $d\leq 3$. 

We note, though, that if the opposite inequality were to hold, $A_H < A_a$,
than $A$ would be negative and $u$ would become complex at large scales.
These two features would indicate a first-order phase transition. This
suggests that the actual first-order phase transition occurs at $t>0$ 
where $A_H$ is effectively smaller, cf. Eq.\ (\ref{eq:4.7c}).

\subsection{Critical behavior}
\label{subsec:IV.B}

\subsubsection{Critical behavior in $d=3$}
\label{subsubsec:IV.B.1}

We are now in a position to determine the critical behavior at the second
order transition that we have shown in the previous subsection to exist in
a certain regime of parameter values. In $d=3$ we do so by
using the explicit solution of the flow equations given in
Sec.\ \ref{subsubsec:IV.A.2} above.  

Let us consider the paramagnon propagator in the critical regime. Since
$t$, $c_1$, and $G$ are not singularly renormalized at one-loop order,
while the coefficient $a$ acquires a nontrivial renormalization, it
has the form, see Eq.\ (\ref{eq:2.9c}),
\bse
\label{eqs:4.13}
\be
{\cal M}_n({\bm k}) = 1/(t + a({\bm k})\,{\bm k}^2 
   + \vert\Omega_n\vert/\vert {\bm k}\vert) \quad.
\label{eq:4.13a}
\ee
${\bm k}$ and $\Omega_n$ have been made dimensionless by means of suitable
microscopic scales. The ${\bm k}$ dependence of $a$ follows from 
Eqs.\ (\ref{eq:4.6}) and (\ref{eqs:4.11}) once $\eta_M$ has been chosen.
\footnote{We remind the reader that the choice of $\eta_M$ is in principle
 arbitrary, although some choices make the analysis of the critical behavior
 easier than others.}
In writing Eq.\ (\ref{eq:4.13a}) we have tacitly assumed that there is no
`wavefunction renormalization' that would give the numerator a scale 
dependence. Since $\eta_M$ determines the scale dimension of the magnetization
field, see Eq.\ (\ref{eq:4.1b}), and hence the wavefunction renormalization,
we need to choose $\eta_M=0$ in order to be consistent with this assumption.
From Eqs.\ (\ref{eq:4.6}) and (\ref{eqs:4.11}) in the limit
$\ell \sim \ln 1/\vert{\bf k}\vert \rightarrow \infty$ we then obtain
\be
a({\bm k}\rightarrow 0) \propto (\ln 1/\vert{\bm k}\vert)^{-1/26}\quad,
\label{eq:4.13b}
\ee
\ese%
Such logarithmic corrections to power-law scaling can be conveniently 
expressed in terms of scale dependent critical exponents. For instance, 
with $b\sim 1/\vert{\bm k}\vert$ a RG length scale factor
\footnote{We use the notation $a\sim b$ for ``scales like $b$'',
and $a\propto b$ for ``$a$ is proportional to $b$''.}
we can write $a({\bm k})\,{\bm k}^2 \propto \vert{\bm k}\vert^{2-\eta}$, 
with a scale dependent critical exponent $\eta$ given by
\bse
\label{eqs:4.14}
\be
\eta = \frac{-1}{26}\,\ln\ln b/\ln b\quad.
\label{eq:4.14a}
\ee
We stress that $\eta$ is the physical critical exponent that describes the
wavenumber dependence of the paramagnon propagator at criticality,
${\cal M} \propto \vert{\bm k}\vert^{-2+\eta}$, as opposed
to $\eta_M$, which has no direct physical meaning. In Appendix \ref{app:A} we
demonstrate that a different choice of $\eta_M$ leads to the same physical
result.

The correlation length exponent $\nu$, the susceptibility exponent $\gamma$,
and the dynamical exponent $z$ can be directly read off Eqs.\ (\ref{eqs:4.13}),
viz.
\be
\nu = 1/(2-\eta)\quad,\quad z=3-\eta\quad,\quad\gamma = 1\quad.
\label{eq:4.14b}
\ee
These exponents are defined as usual, i.e., $\xi\propto t^{-\nu}$,
$\Omega\sim T\sim\xi^{-z}$, ${\cal M}\propto t^{-\gamma}$, with $\xi$ the
correlation length. The physical dynamical
exponent $z$ is different from the exponent $z_M$ in Eq.\ (\ref{eq:4.5b}),
for the same reason for which $\eta \neq \eta_M$. Also notice that $1/\nu$
is not given by the scale dependence of $t$ that results from the $t$-flow
equation, Eq.\ (\ref{eq:4.2a}), since the scale dependent coefficient $a$ is
a dangerous irrelevant variable with respect to the correlation length.

The order parameter exponents $\beta$ and $\delta$ (defined by
$m\propto t^{\beta}$ and $m\propto h^{1/\delta}$, respectively, with 
$h$ a magnetic field) can be obtained from
scaling arguments for the free energy, see Appendix \ref{app:B}. We find
\be
\beta = 1/2\quad,\quad\delta = 3\quad.
\label{eq:4.14c}
\ee

Finally, we define a specific heat exponent $\alpha$ by
$C_{\text{V}}\propto T^{-\alpha}$ at criticality.
\footnote{This is a generalization of the usual definition of $\alpha$ at 
 thermal phase transitions.}
It can be determined by either of three methods, viz., (1) Eq.\ (\ref{eq:4.8})
together with the solution of the flow equation for $H$, (2) renormalized
perturbation theory for the free energy, i.e., Eq.\ (\ref{eq:4.13b}) in 
Eq.\ (\ref{eq:4.9}), or (3) a scaling argument for the free energy, see
Appendix \ref{app:B}. Either way we obtain the exact relation
\be
\alpha = -1 + (\ln\ln b/\ln b - \eta)/z \quad.
\label{eq:4.14d}
\ee
\ese%

The result for $\eta$ is valid to leading logarithmic accuracy; the values of
$\gamma$, $\beta$, and $\delta$, as well as the relations between $\eta$ and
$\nu$, $z$, and $\alpha$, respectively, are exact.

\subsubsection{Critical behavior in $d<3$}
\label{subsubsec:IV.B.2}

In dimensions less than three, the critical behavior can be controlled
by means of an expansion in $\epsilon = 3-d$. We are again looking for
a fixed point where $G$ and $c_1$ are marginal, so Eqs.\ (\ref{eq:4.5a})
and (\ref{eq:4.5b}) still hold. Equation (\ref{eq:4.5c}) gets generalized
to
\be
z_q +\eta_M = 1 + \epsilon\quad,
\label{eq:4.15}
\ee
which guarantees that $c_2$ with $z=z_q$ is still marginal. We then look
for a fixed point where $a$ and $H$ are both marginal. $\eta_M$ then
coincides with the physical exponent $\eta$, as it does in the alternative
treatment of the case $d=3$ given in Appendix \ref{app:A}. We find
\bse
\label{eqs:4.16}
\be
\eta = -\epsilon/(A_H/A_a - 1) = -\epsilon/26 + O(\epsilon^2)\quad.
\label{eq:4.16a}
\ee
The other exponent follow from this. $\nu$, $z$, $\gamma$, $\beta$, and
$\delta$ are still given by Eqs.\ (\ref{eq:4.14b},\ref{eq:4.14c}), and
for the specific heat exponent $\alpha$ we have
\be
\alpha = -d/(3-\eta)\quad.
\label{eq:4.16b}
\ee
\ese%

\section{Summary and Discussion}
\label{sec:V}

We summarize the achievements of this paper as follows.

First, we have given an effective
field theory that describes the quantum ferromagnetic transition in clean
electronic systems. It involves coupled fields describing the magnetization
degrees of freedom, as well as gapless fermionic excitations. If the effects
of the latter are neglected beyond tree level, as was the case in earlier 
theories describing this quantum phase transition,\cite{Hertz_76} then the 
resulting description of the phase
transition is incorrect for all $d\leq 3$. That is, the coupling to the
fermionic degrees of freedom leads to an upper critical dimension for this
phase transition of $d_{c}^{\, +}=3$. 
%It is interesting to note, however,
%that the critical exponents $\beta$, $delta$, and $\gamma$, have their
%mean-field values for all dimensions $d>1$, see Eqs.\ (\ref{eqs:4.14}).
%This shows that $d=1$ is still an upper critical dimension in some sense.
%Only the specific heat, and the wavenumber dependence of the paramagnon
%propagator, which couples to it, are affected by the soft modes that shift
%the upper critical dimension upwards.

Second, we have shown that the fermionic fluctuations lead to two very 
different types of fluctuation-driven quantum phase transitions, depending on
microscopic details. Generically, the quantum ferromagnetic transition in
$d\leq 3$ is a fluctuation-driven first-order transition. This is in contrast 
to the conventional result,\cite{Hertz_76} as well as to the Landau
theory description of this phase transition. In $d=3$ we have also discussed
the situation at low, but finite, temperatures. In general we argue that
this system will have a tricritical point separating lines of second and
first-order phase transitions. A schematic phase diagram is shown in 
Fig.\ \ref{fig:3}. These results are in agreement with the experimental
observations in MnSi\cite{Pfleiderer_et_al_97} and 
UGe$_2$.\cite{Saxena_et_al_00}

Third, we have shown that if the microscopic details are such that the
fluctuation-driven first-order quantum phase transition is too close to a
second-order, or continuous, transition, then critical fluctuations will
suppress the fermionic fluctuation effects that lead to a first-order
transition, and a fluctuation-driven second-order transition results.
\footnote{This second-order transition is unrelated to the one discussed 
 elsewhere,\cite{us_fm_clean} and the two transitions belong to two different
 universality classes. While the one discussed in 
 Ref.\ \onlinecite{us_fm_clean} could be realized somewhere, it is not 
 consistent with low-order perturbation theory, and its
 realization requires something qualitatively to change at higher order. The
 second-order transition discussed in the present paper, on the other hand, 
 is consistent with everything that is known.}
For this case, the critical behavior in $d=3$ has been computed and has been
found to be mean field-like, with logarithmic corrections. For $d<3$, the
critical behavior is nontrivial, but can be controlled by means of a 
$3-\epsilon$ expansion. Both the possibility of a second-order transition
and the fact that the critical behavior in this case is essentially
mean field-like is in agreement with the experimental observations on
ZrZn$_2$\cite{Pfleiderer_et_al_97} and 
Ni$_{\text{x}}$Pd$_{1-\text{x}}$.\cite{Nicklas_et_al_99}
Our theory thus explains a rather confusing experimental situation, where the
transition in bulk systems is observed to be continuous in some systems, and 
discontinuous in others. We further note that the fluctuation effect that leads
to a first-order phase transition grows with the strength of electronic
correlation, or interaction, effects. This suggests that in strongly
correlated systems a first-order transition is generally expected.

Fourth, we have noted a mathematical and physical relation between the
fluctuation-driven first-order phase transition discussed here, and the ones 
known to occur in finite-temperature superconductors and in liquid crystal 
systems.\cite{Halperin_Lubensky_Ma_74,Chen_Lubensky_Nelson_78}
In all these systems, soft modes couple to the order parameter fluctuations
in such a way that their contribution to the free energy is reduced in the
ordered phase. It is this mechanism that causes the discontinuous transition
to occur. The fluctuation-driven second-order transition discussed here is
similar to the one that occurs in classical Potts 
models.\cite{Fucito_Parisi_81}

Elsewhere we have discussed the effects of nonmagnetic disorder on this
phase transition, and on the phase diagram shown in 
Fig.\ \ref{fig:3}.\cite{us_1st_order_PRL} 
In general, sufficiently strong disorder drives the tricritical point shown
in Fig.\ \ref{fig:3} to zero temperature, making the zero-temperature
transition in the presence of sufficiently strong disorder continuous.
This quantum phase transition is in a different universality class than
the fluctuation-driven second-order transition in clean systems discussed
above, and its critical behavior has been determined 
exactly.\cite{us_fm_local_II}

\acknowledgments
Part of this work was performed at the Aspen Center for Physics, and we
thank the Center for its hospitality.
This work was supported by the NSF under grant Nos. DMR-01-32555, 
and DMR-01-32726.

\appendix

\section{RG with scale dependent exponents}
\label{app:A}

Our choice of $\eta_M=0$ in Sec.\ \ref{subsubsec:IV.B.1} is somewhat 
unconventional
since it makes the coefficient $a$ irrelevant, rather than marginal. In this
Appendix we demonstrate that making $a$ marginal leads to the same physical
results, but comes with complications of its own.

Since the physical exponent $\eta$ is scale dependent, see 
Eq.\ (\ref{eq:4.14a}), choosing $a$ marginal requires a scale dependent
$\eta_M$. This changes the flow equations. If we still require that $G$,
$c_1$ (with $z=z_M$) and $c_2$ (with $z=z_q$) are marginal, we obtain,
instead of Eqs.\ (\ref{eqs:4.2}, \ref{eqs:4.7}),
\bse
\label{eqs:A.1}
\bea
\frac{dt}{d\ell} &=& \left(2-{\tilde\eta}_M(\ell)\right)\,t\quad,
\label{eq:A.1a}\\
\frac{da}{d\ell} &=& -{\tilde\eta}_M(\ell)\,a - A_a/H\quad,
\label{eq:A.1b}\\
\frac{dH}{d\ell} &=& {\tilde\eta}_M(\ell)\,H + \frac{A_H}{a+t}\quad,
\label{eq:A.1c}\\
\frac{dc_2}{d\ell} &=& - c_2\quad,
\label{eq:A.1d}\\
\frac{du}{d\ell} &=& -\left(2-{\tilde\eta}_M(\ell)\right)\,u 
                   - A_u\,\frac{c_2^2}{H}\quad.
\label{eq:A.1e}
\eea
\ese%
Here $c_2$ in Eqs.\ (\ref{eq:A.1d}) and (\ref{eq:A.1e}) refers to the 
irrelevant ($z = z_M$) incarnation of $c_2$, and
\be
{\tilde\eta}_M(\ell) = \ell\,\frac{d\eta_M}{d\ell} + \eta_M\quad.
\label{eq:A.2}
\ee
We now look for a fixed point with respect to which $c_2$ and $u$ are
irrelevant, while $a$ is marginal. The latter condition leads to a 
differential equation for $\eta_M$,
\be
\ell\,\frac{d\eta_M}{d\ell} + \eta_M - \frac{1-K}{\ell} = 0
   \quad,
\label{eq:A.3}
\ee
with $K$ from Eq.\ (\ref{eq:4.12b}). In deriving Eq.\ (\ref{eq:A.3}) we
have used the fact that $g = aH$ obeys a flow equation, at $t=0$,
\be
\frac{dg}{d\ell} = A_H - A_a \quad.
\label{eq:A.4}
\ee
Since $a$ is marginal, $\eta_M$ now
represents the physical exponent $\eta$. Eq.\ (\ref{eq:A.3}) is easy to
solve, and for asymptotically large values of $\ell$ one recovers
Eq.\ (\ref{eq:4.14a}). Similarly, the correlation length exponent $\nu$
is now given by the scale dimension of the relevant operator $t$, and
from Eq.\ (\ref{eq:A.1a}) we recover the first equality in 
Eq.\ (\ref{eq:4.14b}). It is easy to check that all other physical results
also agree with Sec.\ \ref{subsubsec:IV.B.1}.

\section{Scaling arguments for the free energy}
\label{app:B}

In this Appendix we consider the scaling behavior of the free energy density.
Let us add a magnetic field term ${\cal A}_h$ to our action,
\be
{\cal A}_h = -{\bm h}\cdot\int dx\ {\bm M}(x)\quad,
\label{eq:B.1}
\ee
${\bm h}$ gets rescaled via ${\bm h} \rightarrow b^{[h]}{\bm h}$
with a scale dimension
\be
[h] = \frac{1}{2}\,\left(d+2+z_M-\eta_M\right)\quad,
\label{eq:B.2}
\ee
and the free energy density obeys the scaling law
\be
f(t,T,h) = b^{-(d+z_M)}\,f(t\,b^{1/\nu},T\,b^{z_M},h\,b^{[h]})
     \quad.
\label{eq:B.3}
\ee
The magnetic susceptibility is given by $\chi = \partial^2f/\partial h^2$,
and it is readily checked that Eq.\ (\ref{eq:B.3}) reproduces the 
susceptibility exponent $\gamma = 1$ that we obtained in 
Sec.\ \ref{subsec:IV.B} from the paramagnon propagator. The critical
behavior of the magnetization, $m = \partial f/\partial h$ can be
obtained from Eq.\ (\ref{eq:B.3}) as well if we take into account that
the irrelevant variable $u$ is a dangerous irrelevant operator with
respect to $m$ (but not with respect to $\chi$). We thus need to include
$u$ in the set of scaling variables, and find
\bse
\label{eqs:B.4}
\be
m(t,h,u) = b^{-(d+z_M-[h])}\,m\left( t\,b^{1/\nu},h\,b^{[h]},u\,b^{[u]}\right)
           \quad,
\label{eq:B.4a}
\ee
with 
\be
[u] = -(d-1+\eta_M)\quad,
\label{eq:B.4b}
\ee
\ese%
the scale dimension of $u$, see Eqs.\ (\ref{eq:4.2c}, \ref{eq:4.5c}). 
Taking into account $m(h=0)\propto u^{-1/2}$ and $m(t=0)\propto u^{-1/3}$,
we obtain mean-field values, Eq.\ (\ref{eq:4.14c}), for the order parameter
exponents $\beta$ and $\delta$.

We also comment on the relation between the free energy scaling and the
specific heat exponent $\alpha$. From Eq.\ (\ref{eq:B.3}) we obtain a scaling
law for the specific heat coefficient $\gamma_V = \partial^2 f/\partial T^2$
at criticality,
\be
\gamma_V (T) = b^{z_M-d}\,\gamma_V(T\,b^z_M)\quad.
\label{eq:B.5}
\ee
In $d=3-\epsilon$ this agrees with the result for the exponent $\alpha$,
Eq.\ (\ref{eq:4.16b}), as obtained from either the $H$-flow equation, or
renormalized perturbation theory. In $d=3$, however, the simple scaling
argument does not agree with the other two methods. The reason lies in
the fact that in $d=3$, $d=z_M$ apart from logarithmic corrections to
scaling. This is one of the `resonances' between exponents that have been
discussed by Wegner in his classification of sources of logarithmic corrections
to scaling.\cite{Wegner_76} This resonance leads to an additional logarithm
that is missed by the simple scaling argument. Once this is taken into account,
all three methods agree in $d=3$ as well.
\bibliography{1st_long}
\end{document}